\documentclass[options]{JHEP3}
\usepackage{epsfig}

\title{Matrix product representation of gauge invariant states 
in a $Z_2$ lattice gauge theory}

\author{Takanori Sugihara\\
RIKEN BNL Research Center, \\
Brookhaven National Laboratory, 
Upton, New York 11973, USA\\ E-mail: \email{sugihara@bnl.gov}}
\abstract{
The Gauss law needs to be imposed on quantum states to guarantee 
gauge invariance when one studies gauge theory in hamiltonian formalism. 
In this work, we propose an efficient variational method 
based on the matrix product ansatz for 
a $Z_2$ lattice gauge theory on a spatial ladder chain. 
Gauge invariant low-lying states are identified by evaluating 
expectation values of the Gauss law operator after numerical 
diagonalization of the gauge hamiltonian. 
}


\preprint{}

\begin{document}

\section{Introduction}

The importance of the first-principle study 
in quantum chromodynamics is increasing largely 
because RHIC experiment has started and LHC is also coming. 
For precise description of high-energy heavy ion collisions, 
gauge theory needs to be studied at finite temperature and density 
in a systematic way. 
Ideally, we should also have a methodology for tracing time-evolution 
of quantum states based on the Schr\"odinger equation 
because heavy ion collisions should be treated as 
non-equilibrium evolving systems rather than static. 
Lattice gauge theory is the most useful method for 
studying the quark-gluon systems at zero and finite temperature. 
However, Monte Carlo integration does not work for 
lattice gauge theory with large chemical potential 
because of the severe sign problem. 
It would be worthwhile to pursue a systematic variational approach 
to gauge theory \cite{Harada:1993va,Sugihara:xq,Sugihara:1997xh,
Sugihara:2001ch,Sugihara:2001ci}. 
In this paper, we apply the matrix product ansatz to a $Z_2$ 
hamiltonian lattice gauge theory on a spatial ladder lattice 
based on the previous work in a U(1) lattice gauge theory 
\cite{Sugihara:2004gx}. 

The matrix product ansatz \cite{or} is a variational method 
that originates in DMRG (density matrix renormalization group) 
\cite{white1,white2,dmrglec}. 
DMRG has been developed as the method that gives the most 
accurate results for spin and fermion chain models 
such as one-dimensional quantum Heisenberg and Hubbard models 
at zero and finite temperature \cite{ft,wx,shiba}.
\footnote{By ``$d$-dimensional'', we mean ($1+d$)-dimensional spacetime.}
DMRG is also useful for 
diagonalization of transfer matrices in two-dimensional classical 
statistical systems \cite{nishino}. DMRG has been extended to 
two-dimensional quantum systems \cite{xiang,farnell} and can work 
for bosonic degrees of freedom \cite{Sugihara:2004qr}. 
Since DMRG is a variational method based on diagonalization of 
hamiltonian and transfer matrices, it is free from the sign problem. 
Actually, DMRG has been successful in accurate study of 
$\theta$-vacuum in the massive Schwinger model 
\cite{Byrnes:2002nv}. 
There is an old prediction by S. Coleman that quarks deconfine at
$\theta =\pi$ \cite{Coleman:1976uz}. 
However, the model has not been analyzed accurately with the 
Monte Carlo method because the topological terms give complex action 
in the Euclidean theory. 

The matrix product ansatz is a result of large simplification 
of trial wavefunction based on the knowledge established in the 
past DMRG studies. 
Therefore, the ansatz takes over the good points of DMRG. 
Besides, the matrix product ansatz has advantage to DMRG because 
the former can treat periodic one-dimensional systems accurately 
\cite{vpc}. 
Recent interesting progress of the matrix product ansatz is 
its application to non-equilibrium quantum physics and 
quantum information theory \cite{schollwoeck}. 
The matrix product ansatz is a promising approach 
to further refinement and extension of the past DMRG studies. 

In general, calculation of expectation values of hamiltonian 
becomes difficult exponentially as the system size increases. 
If the matrix product ansatz is introduced, the energy function 
has a simple matrix product form, which can be evaluated 
easily using a computer if the matrix size is small. 
It is expected that exponentially difficult problems are reduced 
into small tractable ones by the matrix product ansatz. 
Actually, the ansatz has been successful in giving accurate results 
in many one-dimensional quantum systems, 
where calculation errors can be controlled systematically. 
We can say that the matrix product ansatz is 
a first-principle variational method. 

Lattice gauge hamiltonian is obtained by choosing temporal gauge 
in partition function of Euclidean lattice gauge theory 
\cite{Creutz:1976ch}. 
In hamiltonian formalism, gauge invariance needs to be 
maintained explicitly by imposing the Gauss law on the Hilbert space. 
It is a hard task to construct gauge-singlet variational space 
for general gauge group \cite{Robson:1981ws}. 
On the other hand, Euclidean lattice gauge theory can keep 
gauge invariance manifestly by construction.  
This is one of the reasons why hamiltonian version of lattice 
gauge theory is not popular. In addition, no systematic 
methods had been known for diagonalization of gauge hamiltonian 
before the matrix product ansatz was applied to 
lattice gauge theory in ref. \cite{Sugihara:2004gx}. 
If trial wavefunction is constrained directly with the Gauss law, 
the advantage of the matrix product ansatz is completely spoiled 
because calculation of energy function becomes impossible 
in a practical sense. 
If the hamiltonian is diagonalized without the Gauss law, 
all possible states are obtained including gauge variant states. 
However, it must be possible to extract gauge invariant states 
because all eigenstates of the hamiltonian can be classified 
using generators of the considered gauge group. 
Therefore, if the matrix product ansatz is used, 
we better start from the whole Hilbert space and then 
identify gauge invariant states using the Gauss law operator 
after all calculations. 

In hamiltonian lattice gauge theory, the dimension of spatial 
lattice needs to be two or larger because the plaquette operator 
is a two-dimensional object. In this paper, we study a $Z_2$ 
lattice gauge theory on a spatial ladder chain for simplicity.  
The length of the chain needs to be sufficiently long 
for the matrix product ansatz to work well. 
The ladder chain is squashed into a chain 
because one-dimensional structure needs to be found 
in order to use the matrix product ansatz. 

In the previous work \cite{Sugihara:2004gx}, we have studied 
one- and two-dimensional $S=1/2$ Heisenberg models and 
a U(1) lattice gauge theory on a ladder chain 
using the matrix product ansatz in a similar way, 
where energy function is minimized using the Powell method. 
When the number of parameters is very large, 
such naive minimization is not useful because it takes 
long time to reach the bottom of the energy function. 
In this work, we use a diagonalization method introduced in 
ref. \cite{vpc} to obtain sufficient accuracy. 

This paper is organized as follows. 
In section \ref{z2gauge}, hamiltonian lattice formulation 
of the $Z_2$ lattice gauge theory is briefly reviewed. 
In section \ref{mpa}, the matrix product ansatz is introduced 
and applied to a $Z_2$ lattice gauge hamiltonian. 
In the original construction, the matrix product states 
is assumed to have translational invariance. 
In this work, that condition is not imposed on variational space 
before diagonalization of the gauge hamiltonian. 
In section \ref{numerical}, numerical results are given. 
Section \ref{summary} is devoted to summary. 

\section{Quantum hamiltonian in the $Z_2$ lattice gauge theory}
\label{z2gauge}
We are going to introduce the $Z_2$ lattice gauge theory, 
which was invented by F. Wegner \cite{wegner}. 
As seen in the literature, the simplicity of the model is useful 
for testing a new idea 
\cite{Balian:1974ir,Creutz:1979kf}. 
The model cannot have non-vanishing magnetization 
because local gauge symmetry cannot break spontaneously, 
which is known as the Elitzur's theorem \cite{Elitzur:1975imw}. 
However, the model can have nontrivial phases 
depending on dimensionality. 

We are interested in quantum hamiltonian of the model. 
Statistical mechanics and quantum hamiltonian are connected 
through the transfer matrix formalism. 
In the $Z_2$ lattice gauge theory, quantum hamiltonian is obtained 
by choosing temporal gauge 
in the partition function \cite{Kogut:1979wt} 
\begin{equation}
  H =
  -\sum_{n,i} \sigma_x(n,i)
  -\lambda \sum_{n,i,j} P(n,i,j), 
  \label{hamiltonian}
\end{equation}
where $\sigma_x$ and $\sigma_z$ are spin operators 
\[
\sigma_x =
\pmatrix{
  0 & 1\cr
  1 & 0
},\quad
\sigma_z =
\pmatrix{
  1 & 0\cr
  0 & -1
}, 
\]
and $P$ is a plaquette operator 
\begin{equation}
P(n,i,j)\equiv
\sigma_z(n,i)
\sigma_z(n+i,j)
\sigma_z(n+i+j,-i)
\sigma_z(n+j,-i).
\end{equation}
In eq. (\ref{hamiltonian}), 
the first and second summations are taken on the spatial lattice 
for all possible link and plaquette operators, respectively. 
In general, arbitrary states can be represented as a 
superposition of products of $|\pm\rangle_{n,i}$, 
which are eigenstates of the spin operator $\sigma_z(n,i)$
\[
\sigma_z(n,i) |\pm\rangle_{n,i} = \pm |\pm\rangle_{n,i}. 
\]

Let us introduce time-independent operators $G(n)$, each of which 
flips spins on all the links emerging from a site $n$ 
\begin{equation}
  G(n) = \prod_{\pm i}\sigma_x(n,i). 
\end{equation}
We have 
\begin{eqnarray*}
  G^{-1}(n) \sigma_x(m,i) G(n) &=& \sigma_x(m,i), \\
  G^{-1}(n) \sigma_z(n,i) G(n) &=&-\sigma_z(n,i), \\
  G^{-1}(n) \sigma_z(m,i) G(n) &=& \sigma_z(m,i), 
\end{eqnarray*}
where the last formula applies only if the link $(m,i)$ is 
not contained in $G(n)$. The operator $G(n)$ defines 
local gauge transformation 
\begin{equation}
  G(n)^{-1}HG(n) = H. 
  \label{gauge}
\end{equation}
In order for physical quantities to be gauge invariant, 
quantum states need to be invariant under gauge transformation 
\begin{equation}
  G(n)|\Psi\rangle = |\Psi\rangle. 
  \label{gauss}
\end{equation}
We need to impose the Gauss law (\ref{gauss}) on 
the wavefunction to keep gauge invariance. 
Otherwise, unphysical states may be obtained 
because gauge invariance is not guaranteed. 
When a state $|\Psi\rangle$ satisfies the Gauss law (\ref{gauss}), 
magnetization vanishes because a relation 
$\langle\Psi|\sigma_z(n)|\Psi\rangle =
 -\langle\Psi|\sigma_z(n)|\Psi\rangle$ holds. 

\section{Matrix product ansatz on a ladder lattice}
\label{mpa}
We are going to introduce the matrix product ansatz, 
which is a variational method inspired from 
density matrix renormalization group (DMRG). 
DMRG is a variational method that can reproduces 
very accurate results in one-dimensional quantum systems 
\cite{white1,white2,Sugihara:2004qr}. 
In DMRG, wavefunction is represented as a product of orthogonal 
matrices because basis states are rotated for optimization 
with orthogonal matrices that diagonalize density matrices. 
The success of DMRG allows us to parametrize wavefunction 
as a product of finite-dimensional matrices from the beginning. 
This simplification of wavefunction 
is called the matrix product ansatz \cite{or}. 
Although DMRG has slow convergence in one-dimensional quantum 
systems with periodic boundary conditions, 
the matrix product ansatz gives much better accuracy \cite{vpc}. 

\FIGURE{
 \epsfig{file=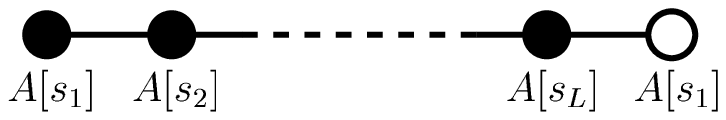,width=9cm}
\caption{A spatial chain with lattice size $L$. 
The open circles indicate periodicity. 
The site variables are dynamical. 
The same set of matrices $A[s]$ is assigned to all the sites. 
}
\label{chain}}
In the original construction, the matrix product state is
parametrized as follows \cite{or}: 
\begin{equation}
  |\Psi\rangle = {\rm tr} \left(\prod_{n=1}^{L}
  \sum_{s_n} A[s_n] |s_n\rangle\right), 
  \label{single}
\end{equation}
where $\{ |s_n\rangle \}$ is a complete set of basis states 
for the $n$-th site (see figure \ref{chain}). 
Hamiltonian needs to have periodicity for consistency with 
the trace operation in the variational state. 
As a result, energy becomes a function of the matrices $A[s]$. 
The minimum of the energy function corresponds to the 
ground state. We are going to apply the ansatz to a $Z_2$ 
gauge theory and see its compatibility with gauge symmetry. 

Since this work is the first application of the matrix product 
ansatz to $Z_2$ gauge theory, 
we would like to consider a simple model. 
The simplest one is a $Z_2$ hamiltonian lattice gauge theory 
on a spatial ladder lattice (see figure \ref{ladder}). 
We assume periodicity in the horizontal direction 
on the ladder for later convenience. In figure \ref{ladder}, 
periodicity is denoted with the open circles. 
\FIGURE{
 \epsfig{file=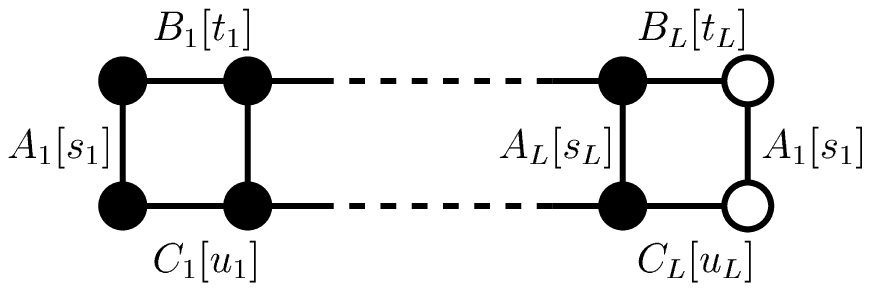,width=11cm}
\caption{A spatial ladder chain with lattice size $L$. 
The open circles indicate periodicity. 
The link variables are dynamical. 
Different sets of matrices are assigned to links. 
}
\label{ladder}
}

In the free case $\lambda=0$, the hamiltonian (\ref{hamiltonian}) 
can be diagonalized analytically. 
The vacuum state is given by 
\begin{equation}
  |{\rm vac}\rangle = \prod_{n,i}
  \frac{1}{\sqrt{2}}(|+\rangle_{n,i} + |-\rangle_{n,i}).
  \label{free}
\end{equation}
Vacuum expectation values of the hamiltonian and plaquette 
operators are $\langle H\rangle /L = -3$ and 
$\langle P \rangle = 0$, respectively. 
In this case, one-dimensional matrices are sufficient 
to represent the vacuum state (\ref{free}) 
\begin{equation}
  A[+]=A[-]=\frac{1}{\sqrt{2}}. 
\end{equation}

When the coupling constant is very large $\lambda \to \infty$, 
the first term of the hamiltonian 
(\ref{hamiltonian}) can be neglected. As a result, 
the hamiltonian has a diagonal form. 
The vacuum state is given by 
\begin{equation}
  |{\rm vac} \rangle =
  \frac{1}{\sqrt{2^{2L+1}}}
  \sum_{m=0}^{2L} G^{(m)}
  \prod_{n,i}|\pm\rangle_{n,i},
  \label{inf}
\end{equation}
where the operator $G^{(m)}$ represents a summation of 
all possible products that are composed of 
$m$ pieces of different Gauss-law operators $G$. 
The operator $G^{(m)}$ has ${}_{2L}C_m$ terms ($G^{(0)}\equiv 1$). 
In eq. (\ref{inf}), the two definitions are identical. 
Then, we have $\langle H\rangle /L \to -\lambda$ and 
$\langle P \rangle \to 1$ in the limit $\lambda\to\infty$. 
The states (\ref{free}) and (\ref{inf}) are both gauge invariant, 
$G(n)|{\rm vac}\rangle = |{\rm vac}\rangle$. 

The $Z_2$ lattice gauge model has only link variables. 
In our construction, each link is assigned a different set of 
matrices $A_n$, $B_n$, and $C_n$ for parametrization of 
wavefunction (see figure \ref{ladder}). 
The index $n$ represents the $n$-th square on the ladder chain 
and runs from $1$ to $L$. 
The dimension of the matrices is $M$. 
Then, our matrix product state is 
\begin{equation}
  |\Psi\rangle =
  {\rm tr}
  \left(
  \prod_{n=1}^L \sum_{s_n=\pm} \sum_{t_n=\pm} \sum_{u_n=\pm}
  A_n[s_n] B_n[t_n] C_n[u_n]\;
  |s_n\rangle_n |t_n\rangle_n |u_n\rangle_n 
  \right),
  \label{mps}
\end{equation}
where the matrices are multiplied in ascending order keeping 
the order of $A_n B_n C_n$, and the basis states 
$|s\rangle_n$, $|t\rangle_n$, and $|u\rangle_n$ 
are eigenstates of the spin operator $\sigma_z$ as before. 
In this expression, the variables $s$, $t$, and $u$ are used 
instead of the index $i$ to denote the position of the links. 
The implementation of the matrix product ansatz means that 
a ladder lattice has been represented as a one-dimensional system 
with non-nearest neighbor interactions. 
Gauge invariance of matrix product states will be discussed 
in the next section. 

If we require orthogonality of optimum basis states 
according to ref. \cite{or}, we have 
\begin{eqnarray}
  \sum_{j=1}^M \sum_{s=\pm} (X_n[s])_{ij} (X_n[s])_{i'j} &=&
  \delta_{ii'}, 
  \label{orthogonality1}
  \\
  \sum_{i=1}^M \sum_{s=\pm} (X_n[s])_{ij} (X_n[s])_{ij'} &=&
  \delta_{jj'}, 
  \label{orthogonality2}
\end{eqnarray}
where $X$ stands for $A,B$, and $C$. 
If these conditions are not imposed, 
norm of the matrix product state (\ref{mps}) may becomes very small, 
which results in numerical instability. 

Energy 
\begin{equation}
  E = \frac{\langle \Psi|H|\Psi\rangle}{\langle\Psi|\Psi\rangle},
  \label{efunc}
\end{equation}
is a function of the matrices $A_n[s]$, $B_n[t]$, and $C_n[u]$. 
The numerator and denominator can be calculated 
by evaluating trace of a product of $3L$ matrices numerically: 
\begin{eqnarray}
  \langle \Psi|H|\Psi\rangle = && -\sum_{n=1}^L
  {\rm tr}
  \left(
  \left(a_n y_n z_n + x_n b_n z_n + x_n y_n c_n \right)
  \prod_{m=n+1 \bmod L}^{n-1+L \bmod L} w_m +
  \right.
  \nonumber
  \\
  && \hspace{0.5cm} +
  \left.
  \lambda\alpha_n \beta_n \gamma_n
  \alpha_{n+1 \bmod L} y_{n+1 \bmod L} z_{n+1 \bmod L}
  \prod_{m=n+2 \bmod L}^{n-1+L \bmod L} w_m
  \right),
  \label{den}
\end{eqnarray}

\begin{equation}
  \langle \Psi|\Psi\rangle =
  {\rm tr}
  \left(\prod_{m=1}^L w_m\right), 
  \label{num}
\end{equation}
where
\begin{eqnarray*}
  &
  \displaystyle
  a_n \equiv \sum_{s,s'}
  (\sigma_x)_{ss'} A_n^*[s] \otimes A_n[s'], \quad
  b_n \equiv \sum_{t,t'}
  (\sigma_x)_{tt'} B_n^*[t] \otimes B_n[t'], \quad
  c_n \equiv \sum_{u,u'}
  (\sigma_x)_{uu'} C_n^*[u] \otimes C_n[u'], 
  &
  \\
  &
  \displaystyle
  \alpha_n \equiv \sum_{s,s'}
  (\sigma_z)_{ss'} A_n^*[s] \otimes A_n[s'], \quad
  \beta_n \equiv \sum_{t,t'}
  (\sigma_z)_{tt'} B_n^*[t] \otimes B_n[t'], \quad
  \gamma_n \equiv \sum_{u,u'}
  (\sigma_z)_{uu'} C_n^*[u] \otimes C_n[u'], 
  &
  \\
  &
  \displaystyle
  x_n \equiv \sum_{s}
  A_n^*[s] \otimes A_n[s], \quad
  y_n \equiv \sum_{t}
  B_n^*[t] \otimes B_n[t], \quad
  z_n \equiv \sum_{u}
  C_n^*[u] \otimes C_n[u], 
  &
  \\
  &
  w_n \equiv x_n y_n z_n. 
  &
\end{eqnarray*}
The dimension of the matrices on the left hand side is $M^2$. 
By the outer product symbol $\otimes$, we mean 
\[
a_{(i,k),(j,l)} = \sum_{s,s'} \sigma_{ss'} A_{ij}[s] A_{kl}[s']. 
\]
The minimum of the energy function (\ref{efunc}) corresponds to the 
ground state, which can be obtained based on matrix diagonalization 
as explained below. 
We can reduce the minimization problem (\ref{efunc}) 
into a generalized eigenvalue problem \cite{vpc} 
\begin{equation}
  v^\dagger \bar{H} v = E v^\dagger N v, 
  \label{gen}
\end{equation}
where $\bar{H}$ and $N$ are $2M^2$ by $2M^2$ matrices. 
To understand what is going here, let us consider 
how energy can be minimized by varying $A_n[s]$ 
when other matrices are fixed. 
Note that equations (\ref{den}) and (\ref{num}) are 
bilinear of the matrix $A_n[s]$
\begin{eqnarray}
  \langle \Psi|H|\Psi\rangle &=& \sum_{i,j,k,l} \sum_{s,t}
  (A_n^*[s])_{ij} \bar{H}_{(i,j,s),(k,l,t)} (A_n[t])_{kl}, 
  \\
  \langle \Psi|\Psi\rangle &=& \sum_{i,j,k,l} \sum_{s}
  (A_n^*[s])_{ij} N_{(i,j,s),(k,l,t)} (A_n[t])_{kl}, 
\end{eqnarray}
where the matrix $N$ is diagonal for the indices $s$ and $t$. 
Once these expressions are obtained and the variational 
parameters $A_n[s]$ are regarded as a vector $v$, 
the minimization problem (\ref{efunc}) reduces to (\ref{gen}). 

There is one more trick that needs to be implemented. 
As explained in equations (\ref{orthogonality1}) and 
(\ref{orthogonality2}), we encounter numerical instability 
on the right hand side in eq. (\ref{gen}) 
if the above procedure is used as it is. 
This is because the matrix $N$ may have very small eigenvalues 
if the matrices $A_n[s]$, $B_n[s]$, and $C_n[s]$ are varied freely. 
For this reason, we need to impose one of the conditions 
(\ref{orthogonality1}) or (\ref{orthogonality2}) on the matrices. 

A matrix can be decomposed into a product of three matrices, 
which is called singular value decomposition. 
For example, if we regard a tensor $A[s]$ 
as a matrix $A_{i,(j,s)}=(A[s])_{ij}=A_{IJ}$, 
singular value decomposition of $A_{IJ}$ is given by 
\begin{equation}
  A_{IJ} = \sum_{K=1}^M U_{IK} D_K V_{KJ}. 
  \label{svd}
\end{equation}
where $D_K$ are the singular values of the matrix $A$ \cite{nr}. 
The matrices  $U$ and $V$ are orthonormal
\begin{equation}
  \sum_{I=1}^M U_{IK}^* U_{IK'} = \delta_{KK'},\quad
    \sum_{J=1}^{2M} V_{KJ}^* V_{K'J} = \delta_{KK'}. 
\end{equation}
The decomposition $A=U'V$, where $U'=UD$ is a square matrix, 
is the key of the trick. 
Consider a part of the matrix product wavefunction 
\begin{equation}
  C[u] A[s] = C'[u] A'[s], 
\end{equation}
where $(A'[s])_{ij}=V_{i,(j,s)}$ and $C'[u]=C[u]U'$. 
If the decomposition (\ref{svd}) is accurate, 
the both representations give the same result. 
Based on this trick, the eigenvalue problem (\ref{gen}) 
is solved successively for all the sets of the matrices 
starting from the right end in figure \ref{ladder}
\begin{equation}
  C_L[u_L] \to B_L[t_L] \to A_L[s_L] \to \cdots \to
  C_1[u_1] \to B_1[t_1] \to A_1[s_1], 
\end{equation}
which we call {\it sweep}. 
In one sweep process, eq. (\ref{gen}) is solved $3L$ times. 
If this sweep process is repeated several times, 
energy of low-lying states converges to some value. 
Calculation error can be controlled systematically 
by increasing the matrix dimension $M$. 
See Appendix \ref{app} for generation of initial matrices.

\section{Numerical results}
\label{numerical}

The matrix product ansatz assumes large lattice. 
Our lattice size $L=500$ is sufficiently large. 
We solve the generalized eigenvalue problem (\ref{gen}) 
using LAPACK \cite{lapack}. 
For steady states, real matrices are sufficient for 
parameterizing the matrix product state (\ref{mps}). 
Convergence of energy needs to be checked for the number of 
sweeps and the matrix dimension $M$. 
Energy density $E/L$ converges in accuracy of five digits or higher 
after two sweeps when the matrix size $M$ is fixed. 
\TABLE{
\begin{tabular}{lllllll}
\hline
$M$ & $E_0/L$ & $E_1/L$ & $E_2/L$ & $E_3/L$ & $E_4/L$ & $E_5/L$\\
\hline\hline
\multicolumn{5}{c}{$\lambda=0.1$} \\
\hline
$2$ &
$\underline{-3.001}$ & $-2.997$ & $-2.997$ &
$-2.997$ & $\underline{-2.993}$ & $\underline{-2.993}$ \\
$3$ &
$\underline{-3.001}$ & $-2.997$ & $-2.997$ &
$-2.997$ & $\underline{-2.994}$ & $\underline{-2.993}$ \\
$4$ &
$\underline{-3.001}$ & $-2.997$ & $-2.997$ &
$-2.997$ & $\underline{-2.997}$ & $\underline{-2.995}$ \\
\hline
\multicolumn{5}{c}{$\lambda=1$} \\
\hline
$2$ &
$\underline{-3.124}$ & $-3.121$ & $-3.121$ &
$-3.118$ & $\underline{-3.114}$ & $\underline{-3.112}$ \\
$3$ &
$\underline{-3.124}$ & $-3.121$ & $-3.121$ &
$-3.118$ & $\underline{-3.114}$ & $\underline{-3.112}$ \\
$4$ &
$\underline{-3.124}$ & $-3.121$ & $-3.121$ &
$-3.118$ & $\underline{-3.114}$ & $\underline{-3.112}$ \\
\hline
\multicolumn{5}{c}{$\lambda=10$} \\
\hline
$2$ &
$\underline{-10.27}$ & $-10.27$ & $-10.27$
& $\underline{-10.27}$ & $-10.23$ & $\underline{-10.23}$ \\
$3$ &
$\underline{-10.27}$ & $-10.27$ & $-10.27$
& $\underline{-10.27}$ & $-10.26$ & $\underline{-10.23}$ \\
$4$ &
$\underline{-10.27}$ & $-10.27$ & $-10.27$
& $\underline{-10.27}$ & $-10.26$ & $\underline{-10.23}$ \\
\hline
\end{tabular}
\caption{Energy density $E/L$ of six low-lying states is listed 
for $\lambda=0.1,1$, and $10$ when lattice size is $L=500$. 
Good convergence of energy is obtained with small $M$. 
}
\label{conv}
}

\FIGURE{
 \epsfig{file=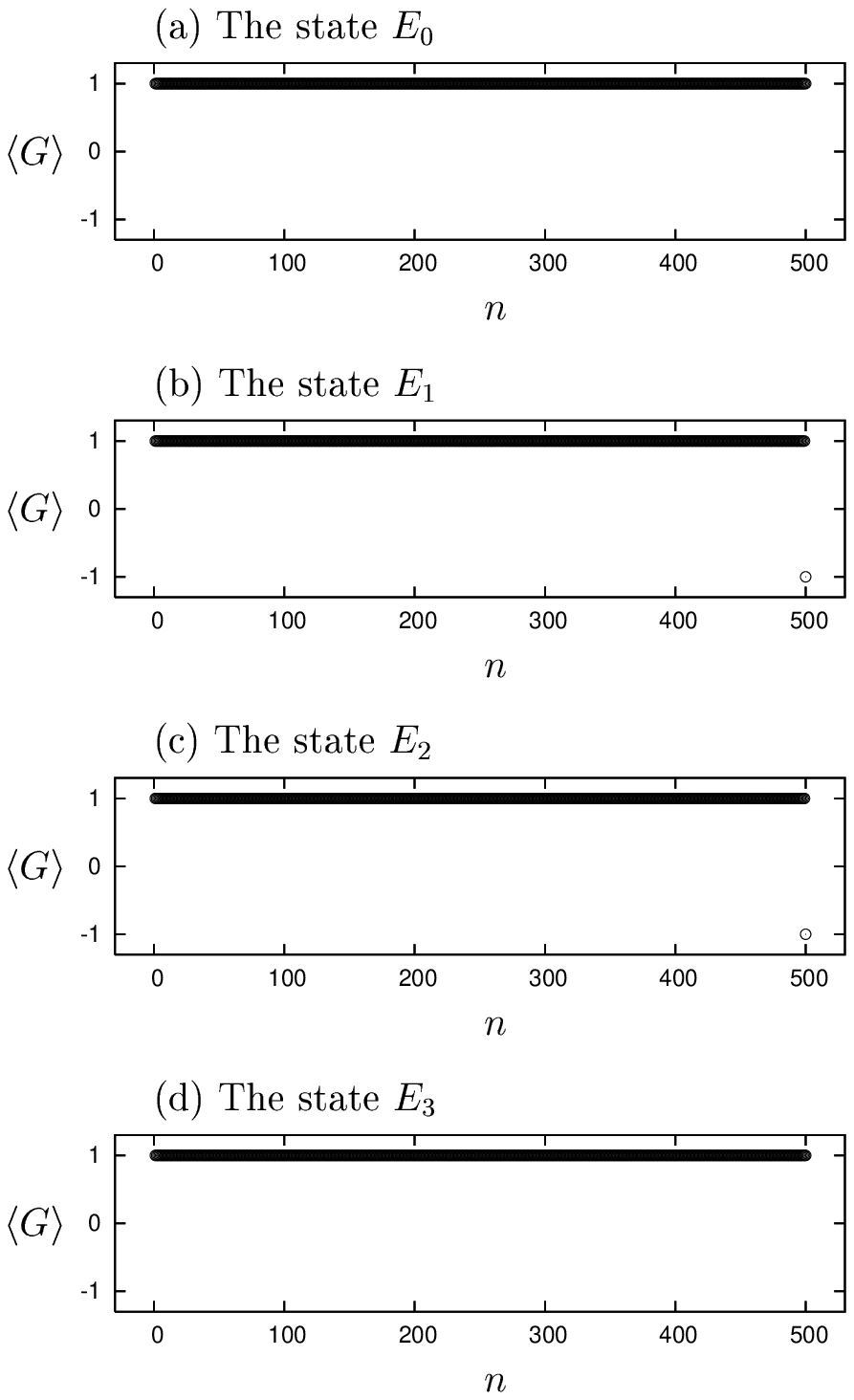,width=9cm}
\caption{Expectation values of the Gauss law operator is 
plotted for the four low-ling states (a) $E_0$, (b) $E_1$, 
(c) $E_2$, and (d) $E_3$ with $\lambda=10$, $L=500$, and $M=4$. 
The circles are the calculated values. 
The states (a) and (d) are gauge invariant because the Gauss law 
is satisfied on every lattice sites. 
On the other hand, the states (b) and (c) are gauge variant 
because $\langle G(500)\rangle =-1$. 
These statements hold in accuracy of seven digits or higher. 
}
\label{gl}}

Table \ref{conv} shows energy spectra of six low-lying states 
for three values of the coupling constant: $\lambda=0.1,1$, and $10$. 
The sweep process has been repeated twice. 
In this model, convergence of energy is very fast 
in contrast to Heisenberg chains \cite{Sugihara:2004gx}\cite{or}. 
Small matrix dimension is sufficient for good convergence. 
Since we have obtained low-lying states without imposing 
the Gauss law on the variational space, 
gauge variant states are contained. 
In table \ref{conv}, gauge invariant states are denoted 
with underlines. The other states are gauge variant. 
As we will see, gauge invariant physical states can be identified 
by calculating expectation values of the Gauss law operator. 

In the ladder chain model, the Gauss law operator $G(n)$ is 
a product of three $\sigma_z$ operators 
(two horizontal and one vertical). 
We evaluate expectation values of $G(n)$ on the upper lattice 
sites shown in figure \ref{ladder}. 
Then, the number of the Gauss law operators to be evaluated is $L$. 
Expectation values on the lower 
sites are same as the upper ones because of reflection symmetry. 
Figures \ref{gl} plots expectation values of the Gauss law operator 
$\langle G(n) \rangle$ in the case of $\lambda=10$ for the states 
(a) $E_0$, (b) $E_1$, (c) $E_2$, and (d) $E_3$. 
In figures \ref{gl} (a) and (d), the Gauss law $G(n)=1$ is satisfied 
uniformly on every lattice sites. Therefore, 
the obtained states $E_0$ and $E_3$ are gauge invariant. 
On the other hand, in figures \ref{gl} (b) and (c), 
the states $E_1$ and $E_2$ are gauge variant 
because gauge symmetry is definitely broken at the site $n=500$. 
The position of this special lattice site depends on 
where the sweep process ends. 
The relation $\langle G(n) \rangle = 1$ or $-1$ holds for 
the obtained low-lying states in accuracy of seven digits or higher 
when $M=4$. 

According to the Elizur's theorem, gauge variant operators 
have vanishing expectation values. 
We have checked that expectation values of single spin operators 
$\sigma_z(n)$ vanish for the gauge invariant states 
in accuracy of ten digits or higher when $M=4$. 
We also have checked that the gauge variant states 
have vanishing expectation values of $\sigma_z(n)$ 
in the same accuracy.  
These statements apply to the low-lying states 
shown in table \ref{conv}. 

In this way, we can classify the obtained states into 
gauge invariant states and others. 
Figure \ref{energy} plots vacuum energy density 
as a function of the coupling constant $\lambda$.

\FIGURE[t]{
 \epsfig{file=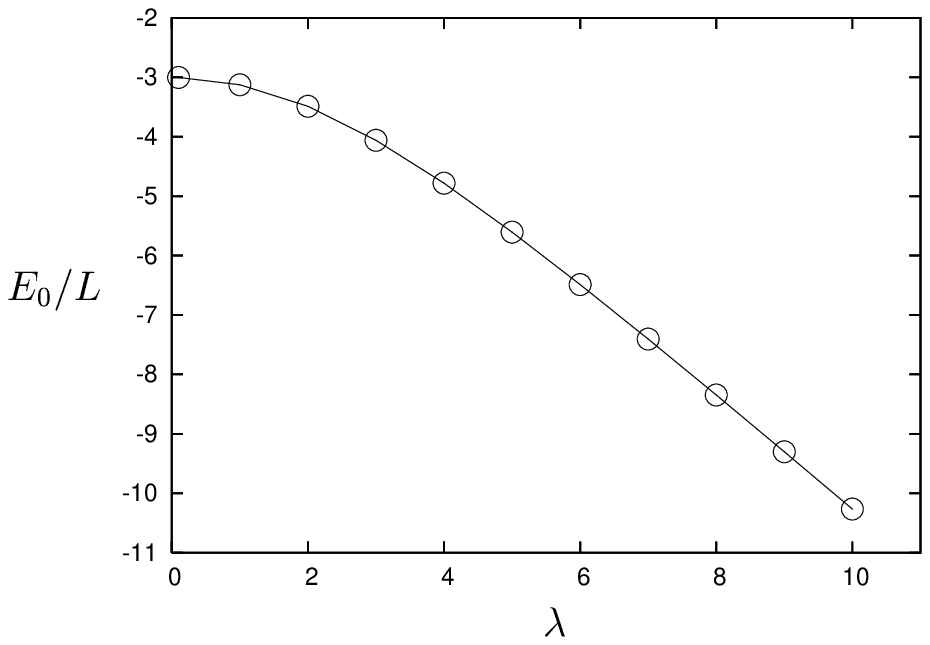,width=9cm}
\caption{Vacuum energy density $E_0/L$ is plotted 
as a function of the coupling constant $\lambda$ 
with lattice size $L=500$. 
The circles are the calculated values, 
which are consistent with the exact ones given in sec. \ref{mpa}. 
}
\label{energy}}
\FIGURE[t]{
 \epsfig{file=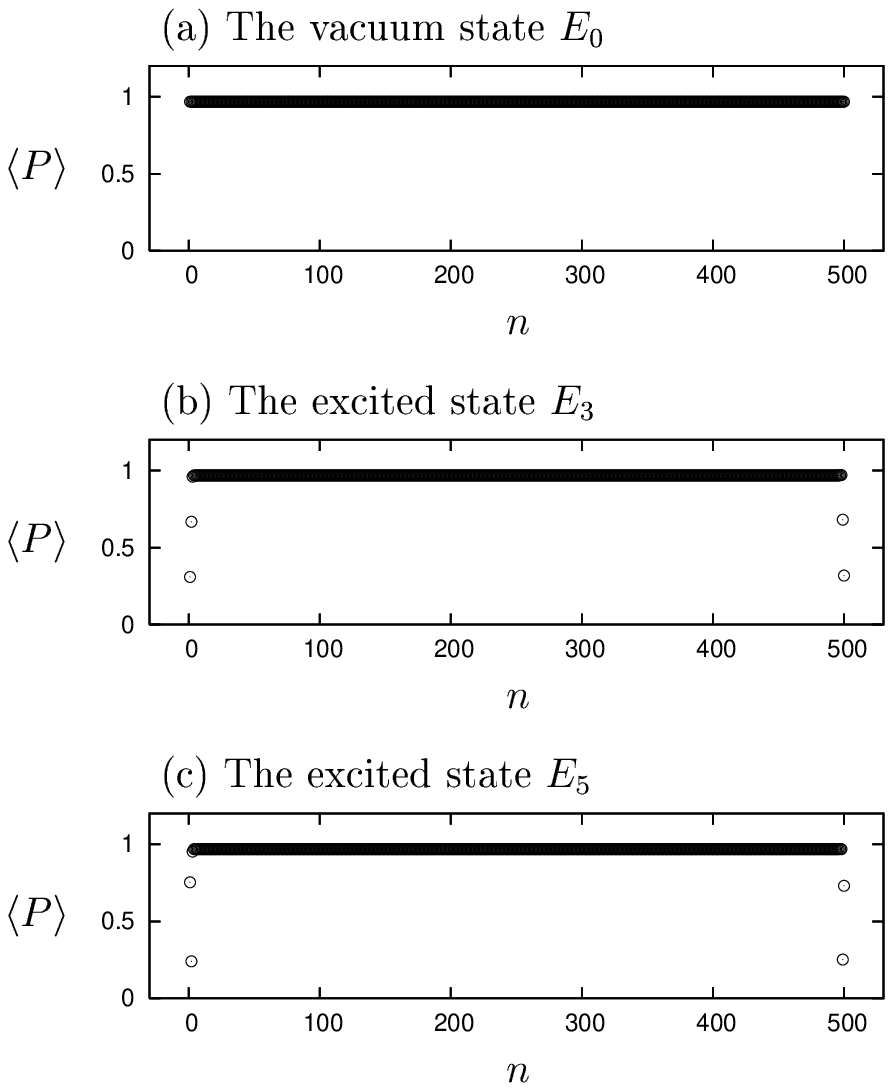,width=9.0cm}
\caption{Vacuum expectation values of the plaquette operator are 
plotted for the gauge invariant states 
(a) $E_0$, (b) $E_3$, and (c) $E_5$ with $\lambda =10$ and $L=500$. 
The excited states (b) and (c) have non-uniformity 
around the boundary $n=500$. 
}
\label{dist}}

In order to see how spatial distribution look like, 
we evaluate expectation values of the 
plaquette operator for the vacuum and the excited states. 
Figure \ref{dist} plots expectation values of the plaquette 
as functions of the spatial lattice coordinate $n$ for the 
gauge invariant low-lying states (a) $E_0$, (b) $E_3$, and (c) $E_5$ 
in the case of $\lambda=10$. 
As expected, the vacuum state $E_0$ has complete uniformity. 
On the other hand, the excited states $E_3$ and $E_5$ have lumps 
around $n=500$. 
The higher excited states have similar lumps around the boundary. 
The obtained solutions have periodicity because the plaquette 
distributions are continuous around the boundary. 
The position of the lumps can be moved without changing energy 
in the sweep process as explained in figure \ref{gl}. 
Figure \ref{plaq} plots vacuum expectation values of the plaquette 
operator as a function of the coupling constant $\lambda$. 
For small and large $\lambda$, the tendency of energy and plaquette 
is consistent with the exact values given in section \ref{mpa}. 

\FIGURE[t]{
 \epsfig{file=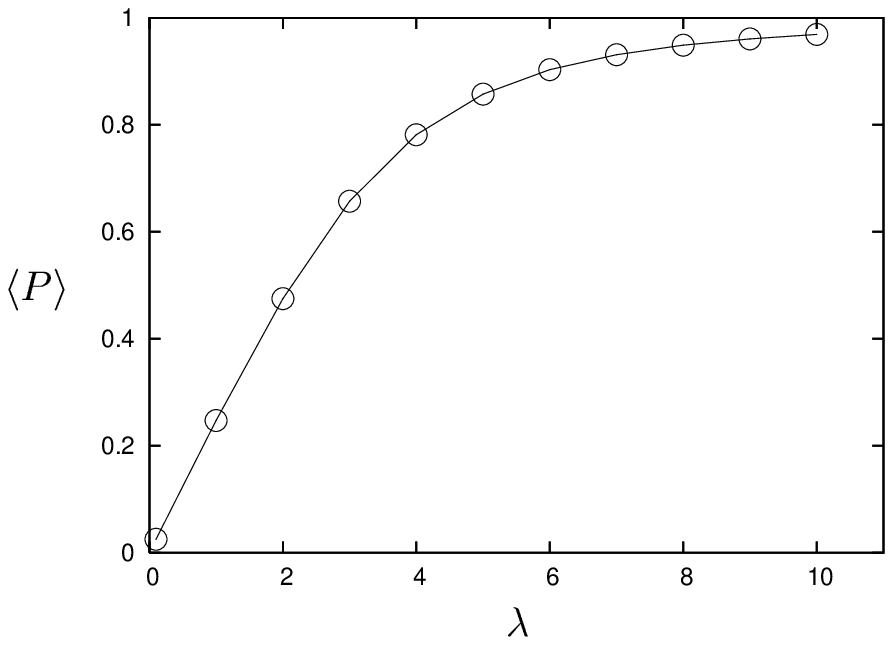,width=9.0cm}
\caption{Vacuum expectation value of the plaquette operator 
is plotted as a function of the coupling constant $\lambda$ 
with lattice size $L$=500. The circles are the calculated values, 
which are consistent with the exact ones given in sec. \ref{mpa}. 
}
\label{plaq}}

\section{Summary}
\label{summary}
We have extracted gauge invariant physical states in a $Z_2$ 
hamiltonian lattice gauge theory on a spatial ladder chain. 
The calculations are based on the matrix product ansatz, 
which gives sufficiently convergent energy and wavefunction 
for low-lying states. 
In the future studies, similar calculations should be tested in 
higher dimensional lattice gauge theory including supersymmetric 
cases \cite{Sugino:2004uv}. 
The proposed method will be useful especially for non-perturbative 
analysis of vacuum structure at the amplitude level.

\acknowledgments
The author would like to thank T.~Nishino for useful communications. 
The numerical calculations were carried on the RIKEN RSCC system. 
This work has been partially supported by RIKEN BNL.

\appendix
\section{Generation of initial matrices}
\label{app}

Before starting sweep process, we need to prepare initial values 
of matrices for the variational state (\ref{mps}). 
As explained in section \ref{mpa}, the matrices need to satisfy 
one of the orthogonality conditions (\ref{orthogonality1}) or 
(\ref{orthogonality2}) for numerical stability. 
The algorithm shown below is also useful for minimization of 
energy function with the Powell method \cite{Sugihara:2004gx}. 

Let us consider real matrices $A[s]$ that 
satisfy the following normalization condition 
\begin{equation}
  \sum_{j=1}^M \sum_{s=1}^K
    A[s]_{ij} A[s]_{i'j}=\delta_{ii'}, 
    \label{orthonormality}
\end{equation}
where $M$ is the dimension of the matrices $A[s]$ and $K$ is the 
degrees of freedom of each site or link. 
To parametrize the matrices $A[s]$, 
we introduce $KM$-dimensional vectors $v^{(n)}$ 
\begin{equation}
  v^{(n)} =
  (v_1^{(n)},\dots,v_{KM-n+1}^{(n)},\underbrace{0,\dots,0}_{n-1}), 
  \quad
  n=1,\dots,M. 
\end{equation}
These vectors are linearly independent and can be
orthonormalized using the Gram-Schmidt method 
\begin{eqnarray}
  a^{(1)} &=& \frac{v^{(1)}}{|v^{(1)}|}
  \label{normalization1}
  \\
  b^{(k+1)} &=& v^{(k+1)}
    -\sum_{n=1}^k \langle v^{(k+1)}, a^{(n)} \rangle a^{(n)}
  \\
  a^{(k+1)} &=& \frac{b^{(k+1)}}{|b_{(k+1)}|}
  \label{normalization2}
\end{eqnarray}
where $k=1,\dots,M-1$ and the brackets represent inner product.
The orthonormalized vectors $a^{(n)}$ are used to 
parametrize the matrices $A[s]_{ij}=A_{i, (j,s)}$
\begin{equation}
  A_{i, (j,s)}=
  \left(
  \begin{array}{c}
    a^{(1)} \\
    \vdots  \\
    a^{(M)} \\
  \end{array}
  \right),
\end{equation}
which satisfy the conditions (\ref{orthonormality}). 
When eq. (\ref{orthonormality}) is satisfied, 
the number of independent degrees 
of freedom associated with the matrices $A[s]$ is given by 
\begin{equation}
  K M^2 - \frac{M(M+1)}{2},
\end{equation}
which is equal to that of the vectors $v^{(n)}$ 
\begin{equation}
  \sum_{n=0}^{M-1} (KM-n) - M. 
  \label{v}
\end{equation}
In eq. (\ref{v}), the first term counts 
the number of the parameters $v_i^{(n)}$ and 
the second term comes from the normalization conditions 
(\ref{normalization1}) and (\ref{normalization2}).

\end{document}